# Effect of entanglement and crosslinking on the hyperelastic behavior of SBR rubber: A multiscale DPD simulation study

Shashank Mishra[a], Soumyadipta Maiti[a,1], Prashant Pandey[b], Beena Rai[a]

[a]TCS Research, Tata Research Development and Design Center, 54-B Hadapsar Industrial Estate, Hadapsar, Pune, 411 013, Maharashtra, India

[b] School of Materials Science and Technology, Indian Institute of Technology (Banaras Hindu University), Varanasi 221 005, Uttar Pradesh, India

[1] Corresponding author email: soumya.maiti@tcs.com

**Abstract**

In this study we have investigated into the entanglement effect and crosslinking effect in long-chain SBR rubber polymer system to model the mechanical hyperelastic behavior. The discussed methodologies are developed by a dissipative particle dynamics (DPD) based multiscale modeling method. The DPD interaction parameters are in turn obtained by all atomistic molecular dynamics and subsequently utilized by the DPD simulations. In the DPD simulation boxes, 200 long polymer chains and 1600 vulcanizing sulphur beads were packed. Vulcanization was achieved by random cross-linking among chains and entanglement was detected by utilizing the M-coil estimator. The mechanical effect of entanglement was modelled by extended tube model and the crosslinking effect was modelled by the Arruda-Boyce model. The modelled stress-strain curve is compared with the experimentally obtained curve for the qualitative and quantitative aspects of mechanical properties. It was determined that for strains up to 150%, the contribution of entanglement to the measured mechanical properties can be even up to 50%.

## 1. Introduction

Polymeric materials are essential for many applications in current times such as light structural components, elastomers, viscoelastic materials etc for various advanced engineering and day-to day applications [1-3]. These polymers generally comprise of different chemical species as their monomers, which are combined to form the long chain of molecules [1, 2]. These long chains of molecules are sometimes crosslinked to improve their properties such as elastic modulus and hyperelastic stretching, yield strength, glass transition temperature etc [1-3]. The network thus formed plays a critical role on the mechanical properties of the polymers [1,2]. These crosslinked networks in elastomers consists of mainly two kind of structural points: entanglement points and crosslinking points which constraints the movement of these chains [1]. These elastomers possess significant hyperelastic properties with the stretchability of the segments of chains between two crosslink points [1]. In recent years, numerical simulation and modelling of elastomer's mechanical behaviour through various constitutive analytical models (CAM) have gained significant importance to study the hyperelasticity with respect to the stress and internal energies of elastomers [1-5]. These constitutive models mainly use a strain-energy functional to describe the stress-strain response in elastomers [1, 3].

In these CAMs, certain parameters that are required are usually derived by fitting the constitutive model to the available experimental data for the imposed deformation states. Alternatively, those can be obtained by the structural analysis of the polymer chains obtained from molecular dynamics (MD) simulations for a given elastomer. These simulations incorporate molecular phenomena and hence they provide more accurate insights into the structural properties of these polymeric systems [1]. MD methods have been successfully applied in the past for prediction of various properties like glass transition temperature (GTT), free volume calculations of elastomers [2] etc. A few hierarchical multiscale approaches were also established in the literature to characterize the mechanical properties of elastomeric systems [3-5]. There are also studies where people have coupled the MD simulations with FEM simulations to obtain certain properties in polymers [4,6,7]. However, there are several computational challenges and complexities associated with simulating bulk behaviour of large systems.

Recently, Chaube et al [8] developed a multiscale model to predict the hyperelastic behaviour in SBR1502 rubber where they coupled the CAM model with all atomistic MD simulations. They were able to predict the hyperelastic model behaviour to a reasonably good accuracy level. However, their model ignored the effect of entanglements in their CAM model and also, they had very limited number of polymeric chains in their MD model to adequately capture the entanglement effects. In this study, we will include the entanglement effects in addition to their model and also we will take larger system. Since all-atomistic MD simulations of larger systems require larger computational power, mesoscale simulation techniques are needed to simulate such large system. In the present study, DPD simulations were used to investigate the structures of these elastomeric systems. There had been various studies on entangled polymer melts through DPD simulations [9,10,11,12] and hence it indicates that DPD can be used to study our system as well.

In this study, we have combined two different constitutive analytical models to investigate the hyperelastic behaviour of our elastomeric system. The elastomer in this study is SBR1502, crosslinked with sulphur crosslinking agents and their entangled and crosslinked structure were simulated and analysed using DPD simulations. We have also discussed the generation of parameters for these DPD simulations through appropriate all-atomistic MD simulations. This paper is organised as following: Section 2 describes the constitutive model we have used, Section 3 describes the simulation methodology which included information of DPD simulation, coarse-graining scheme and also the method to estimate the parameters required for DPD simulation and Section 4 describes the results and discussion of this study.

## 2. Constitutive model description:

The mechanical properties in an unfilled elastomer system is governed by the deformation of the network of the polymeric chains in the system. At the molecular level, these networks consist of entanglement points or points at which polymer chains constrain the movement of other chains and separate crosslinking points created due to vulcanisation process. Hence, when deformed the movement constraints of these specific points govern the mechanical properties. In constitutive analytical models (CAM) from the literature, various models have been tried to link these molecular structure properties with their mechanical properties. Some example of such CAM models include the Mooney-Rivlin[13], Treloar [14], Arruda-Boyce [15], Beda [16], Gent-Gent model [17], Extended Tube model [18], and Ogden model[19].

In this study, some polymer chain network related parameters were estimated using molecular simulations and these parameters were then passed to such CAM models. These parameters do not depend on the fitting of the experimental stress-strain curves. Rather, these models depend on such parameters which can be directly estimated by investigating the molecular topological network of such systems. One such model is the Arruda-Boyce model [15] where the nominal engineering stress-stretch relations can be expressed as:

$$\sigma_{uniaxial} = (\lambda - \lambda^{-2})\frac{NkT\sqrt{n}}{3\lambda_{chain}}L^{-1}\sqrt{\left(\frac{\lambda^2+2\lambda^{-1}}{3n}\right)} \qquad (1)$$

With

$$\lambda_{chain} = \sqrt{\frac{\lambda_1^2+\lambda_2^2+\lambda_3^2}{3}} \qquad (2)$$

This model require two lower length scale parameters – $n$ (average number of random links per chain segment between two cross-linking points) and $N$ (volumetric number density of chain segments) – to predict the deformed network dynamics of polymers [16, 20].

$L^{-1}$ is the inverse Langevin function defined by:

$$L(\beta) = cot(\beta) - \frac{1}{\beta} \qquad (3)$$

The parameters *n* and *N* can be estimated through MD simulations and then the stress-stretch relationship can be predicted as shown in Ref [8], where the authors have predicted the hyperelastic stress-strain curves for SBR 1502 in similar manner. However, the Arruda Boyce model neglect the effect of entanglements on the stress-stretch relationships for elastomers [21]. Thus, this model underestimated the stress levels at lower strain values as it can be seen from Fig. 7 of Ref [8]. At lower strain levels the effect of entanglements is dominant over the effect of crosslinking on the mechanical properties of elastomers, hence it can't be neglected. In this study, to capture this entanglement effect we used a CAM model for entanglement and then integrated this with the above Arruda-Boyce model to capture the uniaxial stress-strain behaviour of the studied elastomer system.

Ref [22] compared the various CAM models and showed that the extended tube model [18] gives excellent quantitative agreement with experiments and ranked it as the best model because it involves only four parameters and its derivation is physically motivated. In the extended tube model (ETM), the equation for uniaxial engineering stress-stretch relationship is given as [18, 23]

$$\sigma_{uniaxial} = G_c(\lambda - \lambda^2)\left\{\frac{1-a}{1-a(I_1-3)^2} - \frac{a}{1-a(I_1-3)}\right\} + 2G_e\left(\lambda^{-1/2} - \lambda^{-2}\right) \qquad (4)$$

where $G_c = \nu_c k_B T$ is the cross-link modulus, $\nu_c$ is the number density of mechanically active network strands between the junctions, $G_e = 0.5\nu_e k_B T$ is the entanglement modulus, $\nu_e$ is the number density of chain fragments between entanglements (entanglement strand), and parameter *a* takes into account the finite extensibility of polymer chains between the network knots. $I_1$ in eq 4 is the first scalar invariant of the Cauchy− Green deformation tensor; it is equal to $I_1 = \lambda^2 + 2/\lambda$ in the case of uniaxial deformation of incompressible solid [24]. Saphiannikova et al [23] tried to fit the experimental stress strain curves of S-SBR at different crosslinking densities to get the values of parameters $G_c$, $G_e$ and $a$ using the above equation (4) to predict the mechanical behaviour of this system. It can be seen from equation (4) that the extended tube model considers the effect of both crosslinking and entanglements added together independently in the stress-strain behaviour of elastomers. However, the parameter $a$ cannot be easily predicted from the molecular topological network of elastomers and mostly estimated through fitting of experimental dynamic mechanical experiments. This implies the effect of crosslinking cannot be easily studied through topological network analysis. Whereas, the entanglement effect is captured by single parameter $G_e$ which in turn is predicted by the parameter $v_e$. This parameter $v_e$ can be estimated by analysing the molecular backbone structures of the elastomers and their topological constraints. In Ref. [23] it was experimentally established that the crosslinking effect and entanglement effects work independently in hyperelastic rubbers and variation in crosslinking density does not affect the entanglement. Hence, we use the entanglement effect of ETM model along with the Arruda-Boyce model (it captures the crosslinking effect reasonably well) to predict the stress-strain relationship in elastomers. Therefore, in this study, the uniaxial stress strain is expressed as following equation:

$$\sigma_{uniaxial} = (\lambda - \lambda^{-2})\frac{NkT\sqrt{n}}{3\lambda_{chain}}L^{-1}\sqrt{\left(\frac{\lambda^2+2\lambda^{-1}}{3n}\right)} + 2G_e(\lambda^{-1/2} - \lambda^{-2}) \quad (5)$$

The temperature $T$ of the system is taken as 300 K for all the analysis.

## 3. Simulation Methodology

In this work, DPD simulations on a system comprising of coarse-grained models of SBR-1502 chains and sulphur molecules as crosslinking agents were carried out. The coarse graining mechanism and the DPD simulation methodology has been discussed in subsequent sections.

### 3.1. DPD Simulation

In the usual model for DPD polymers, the force acting between beads $i$ and $j$ is computed as the sum of a non-bonded conservative force $F_C$, a dissipative force $F_D$, a random force $F_R$, bond force $F_B$ and angle force $F_A$ [25]. In DPD simulation we integrate the following Newton's equations of motion

$$\frac{dr_i}{dt} = v_i, \frac{dv_i}{dt} = f_i \quad (6)$$

Where the total force $f_i$ is denoted by following equation.

$$\boldsymbol{f}_i = \sum_{j \neq i} \boldsymbol{f}_{ij} = \sum_{j \neq i}(\boldsymbol{f}_{ij}^C + \boldsymbol{f}_{ij}^D + \boldsymbol{f}_{ij}^R) + (\boldsymbol{f}_i^B + \boldsymbol{f}_i^A) \quad (7)$$

The conservative force is represented as:

$$f_{ij}^C = a_{ij}(1 - r_{ij})\hat{r}_{ij}, \qquad (r_{ij} < 1) \tag{8}$$

while the dissipative force is represented as:

$$f_{ij}^D = -\gamma w^D(r_{ij})(\hat{r}_{ij} \cdot v_{ij})\hat{r}_{ij} \tag{9}$$

The random force is described as:

$$f_{ij}^R = \sigma w^R(r_{ij})\theta_{ij}\hat{r}_{ij} \tag{10}$$

In DPD simulation following relation between the coefficients always hold true:

$$w^D(r) = [w^R(r)]^2 \tag{11}$$

$$\sigma^2 = 2\gamma k_B T \tag{12}$$

For bonded interactions, harmonic bond and angle potentials were used.

$$E_{bond} = K_b(r - r_0)^2 \tag{13}$$

$$E_{angle} = K_a(\theta - \theta_0)^2 \tag{14}$$

### 3.2. Coarse Graining Scheme:

There are 4 different kinds of monomers that constitute a single SBR1502 chain: styrene, 1,2-butadiene, cis-butadiene and trans-butadiene. These individual monomers have to be represented using beads in the DPD simulations. Typically, in a DPD simulation, the mass and volume of beads of different species should be comparable [26]. Hence, based on the molecular weights of these monomers a mapping criterion was used such that styrene was represented using two beads and other monomers as a single bead because the monomer weight of styrene is approximately double of the other monomers. This is depicted in Fig. 1 for simplification. Each crosslinker molecule was also modelled as a single bead. The beads were distributed in a chain of SBR1502 according to the weight fraction given in the Refs [8,27]. The mole fraction of each bead in a chain of SBR1502 is given in Table 1. Then, we run some all atomistic MD simulation in the NPT ensemble for 1 ns with a timestep of 1 fs to estimate the densities of pure monomer systems at temperature 300 K. The inital configuration were created from Ligpargen [28]. OPLS-AA forcefield parameters were used for the simulations. The estimated densities from these simulations has been tabulated also in Table 1. Using the densities of individual monomers and mole fraction of each monomer beads in a SBR 1502 chains, an average density of a bead was calculated, which came out to be 0.668 g/cc. Also, we calculated the mean molar mass of the beads from the molecular weights of each bead and the atomic fraction of these beads. This mean molar molecular mass of each bead was calculated to be equal to 53.60 g/mol. From the mean molar mass and average density, we can obtain an average volume of the DPD bead, $v$ equal to 133.107 Angs$^3$ (53.60/(Avagadros'number*0.668)).

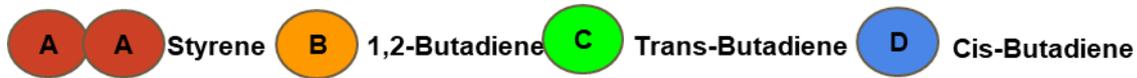

Figure 1. Beads used in DPD simulations

Table 1. Details of beads in the SBR-1502 chain

| Bead Type | A | B | C | D |
|---|---|---|---|---|
| Monomer | Styrene | 1,2-butadiene | Trans-butadiene | Cis-butadiene |
| Molecular mass(g/mol) | 104.15 | 54.096 | 54.096 | 54.096 |
| Mass of each bead (M) | 52.075 | 54.096 | 54.096 | 54.096 |
| Mole Fraction | 0.138 | 0.135 | 0.62 | 0.11 |
| Density of monomer (g/cc) | 0.927 | 0.624 | 0.625 | 0.63 |
| Mole Fraction of Beads | 0.242 | 0.119 | 0.546 | 0.094 |
| No of bead in a chain | 206 | 101 | 465 | 80 |

In a DPD simulations, the reduced bead number density is chosen as $\rho_n = 3$. Thus, following the methods of DPD simulation in Refs [26,29], we can define the length scale of the simulation as $L=(v\rho_n)^{1/3} = 7.36$ Ang. This length scales denotes there are 3 DPD beads in small cell of size $L^3$. Using the information from Table 1, a coarse-grained chain of SBR-1502 comprising of 852 beads was made having molecular weight of 45675 g/mol. The monomers were distributed randomly in the chain.

### 3.3 Calculation of repulsive parameter '$a$'

The DPD interaction parameters between different components were estimated from Flory Huggins parameter ($\chi$) using the method established by Groot and Warren [25] and used by Liu et al [29] for modelling the curing reaction and network structure in an epoxy system. For getting the Flory Huggins parameter ($\chi$) at a specific temperature, MD simulations at those temperatures were run to estimate their solubility parameters. Initially an NVT simulation was run for 100 ps, followed by NPT simulations for 1 ns to get their equilibrium density. Then a final NVT simulation was performed for 1 ns, in which the last 200 ps trajectories were used for calculating ensemble averages. Just like the method followed by Ref [29], for the system containing k molecules of the same species, we can obtain the nonbonded energy of the model, $E_{nb}^{k}$. Then each molecule is extracted to vacuum to calculate the nonbonded energy for the individual molecule. Thus, the cohesive energy $E_{coh}$ can be calculated via the following relation [30]

$$E_{coh} = (\sum_{i=1}^{k} E_{nb}^{isolated}(i) - E_{nb}^{k})/k \quad (15)$$

where, $E_{nb}^{isolated}(i)$ is the nonbonded energy for the ith isolated molecule in vacuum. The cohesive energy densities of the monomers at the experimental temperature, $e_{coh}$, can therefore be calculated as, $e_{coh} = E_{coh}/V_m$, where $V_m$ is the molar volume of the monomer. Subsequently, the solubility parameters of the components can be estimated with the relation $\delta = (e_{coh})^{1/2}$. The simulation results are as shown in Table 2. These results were compared with the experimental results [27] and there was a good agreement between the values. For sulphur the solubility parameter was taken from Ref [31].

Table 2. MD simulation results and experimental values of solubility parameters of monomers and crosslinkers

| Molecule | $E_{coh}$ (kcal/mol) | $\delta(cal/cm^3)^{1/2}$ | Reference Value of $\delta(cal/cm^3)^{1/2}$ |
|---|---|---|---|
| Styrene | 13.13 | 10.8 | 9.15 |
| 1,2-Butadiene | 10.81 | 11.16 | - |
| Trans-butadiene | 55.04 | 7.97 | 8.33 |
| Cis-butadiene | 58.05 | 8.22 | 8.33 |
| Sulphur | - | - | 16.1 |

The Flory-Huggins interaction parameters $\chi$ at 300 K can therefore be calculated via the following relation [32-34]

$$\chi_{ij} = \frac{(\delta_i - \delta_j)^2 v_{ref}}{k_B T} \quad (16)$$

where $\delta_i$ and $\delta_j$ are the solubility parameters for component $i$ and $j$, respectively, and $v_{ref}$ is the reference bead volume. The reference bead volume $v_{ref}$ can be taken as the average size of the coarse-grained beads, which is $v = 133.107$ Angs$^3$ in this study. Therefore, the values of $\chi_{ij}$ between different components are obtained. The empirical relation between Flory-Huggins $\chi$ parameters and DPD interaction parameters is used to evaluate the DPD interaction parameters between different species [25]:

$$\chi = (0.286 \pm 0.002)(\alpha_{ij} - \alpha_{ii}) , (\rho = 3) \qquad (17)$$

where $\alpha_{ii}$ ( = 25) and $\alpha_{ij}$ are the DPD interaction parameters between the similar and dissimilar types of species, respectively. The calculated DPD interaction parameters between different components, $\alpha_{ij}$, are shown in Table 3.

Table 3. Repulsive parameter "$a$" between different DPD beads

| $\alpha_{ij}$ | Styrene | 1, 2-Butadiene | Trans-butadiene | Cis-butadiene | Sulphur |
|---|---|---|---|---|---|
| **Styrene** | 25.000 | 25.060 | 29.053 | 28.384 | 26.835 |
| **1,2-Butadiene** | | 25.000 | 30.136 | 29.379 | 26.213 |
| **Trans-butadiene** | | | 25.000 | 25.030 | 36.341 |
| **Cis-butadiene** | | | | 25.000 | 35.202 |
| **Sulphur** | | | | | 25.000 |

### 3.4. Simulation Methodology

Once all the essential parameters required for DPD simulations were obtained, we started the process of simulating the system using DPD methodology. An amorphous box containing 200 randomly oriented chains of SBR-1502 was created using the method as followed in section 3.1 and 1600 crosslinker beads were also dispersed in the box using PACKMOL [35] with a number density of all beads equal to 3. This system corresponded to rubber systems containing nearly 1.75 parts per hundred rubber (phr) in the blend. The box size was 38.56 x 38.56 x 38.56 in reduced units. Each of the polymer chain molecules were at first made by arranging the bonded beads in a self-avoiding random walk configuration before being packed by PACKMOL. Same methodology as above was used to create 2 other amorphous boxes with different initial random positions of the chains and the crosslinker beads but having same composition and bead density.

Initially, these systems were minimised for energy and then the DPD simulations were carried out for 200000 steps with a timestep of 0.01 in reduced units. Then a crosslinking reaction between the SBR molecules and the crosslinker molecules was carried out during the simulation for another 300000 steps. This reaction carried out in this simulation was similar to the crosslinking reaction given in Ref [8]. A cut-off of 0.9 was used to create the crosslinked bonds. Then further DPD simulation were carried out for another 200000 steps to equilibrate the system. Finally, an additional DPD run was carried out for 800000 steps and then these structures were analysed. A segmental repulsive potential (SRP) was also used during the entire simulation in order to avoid any polymer-chain crossovers [36]. The temperature was taken as 1 in reduced unit throughout the DPD run corresponding to temperature of 300K and energy in reduced unit becomes $300k_B$, where $k_B$ is the Boltzmann constant [37]. Table 4 shows the conversion of units from reduced units to real units. All the simulations were carried out in open source LAMMPS package [38]. Fig. 2 depicts an entangled polymer system obtained after the DPD simulations. The figure depicts a dense structure within the simulation box with differently colored polymer chains dispersed among each other. The value of the constants for the bonded potentials (eq. 13 and eq. 14) like $K_b$, $K_a$, $r_0$ and $\theta_0$ in our DPD simulations were taken as 100, 100, 0.85 and 120 respectively, in reduced units according to Ref [31]. Fig. 3 shows some of the examples of entanglement points with differently colored polymer chains that were found after visually inspecting the system. Entanglement points created by one polymer chain of a particular color restricting the movement of other adjacent chain is indicated by the turns and coils as shown in the cases in Fig. 3.

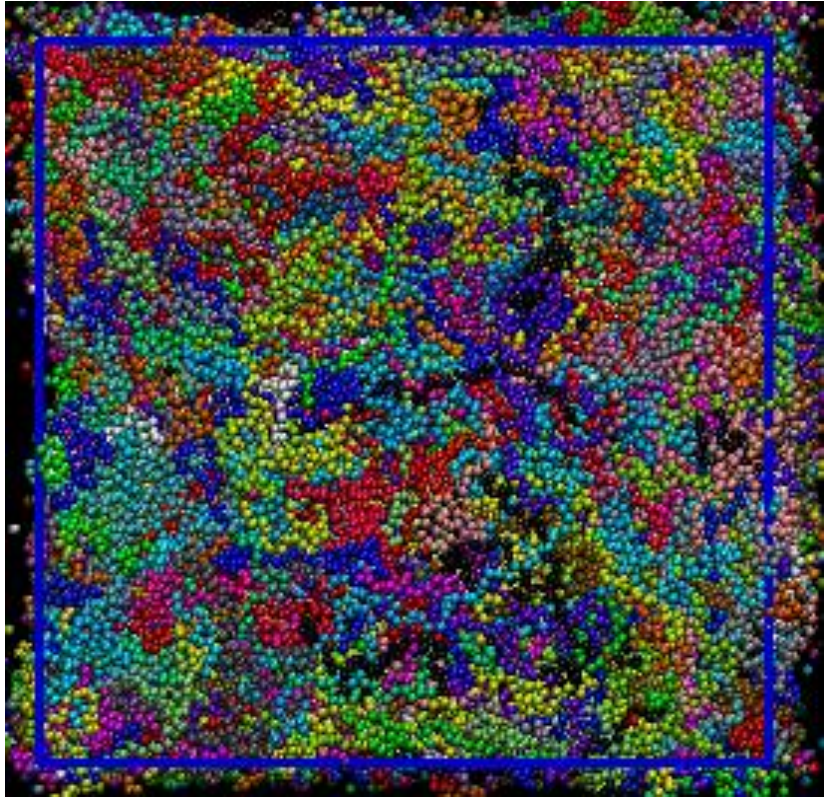

Figure 2. Entangled polymer system obtained after DPD simulation. Here each polymer chain is depicted with different colors.

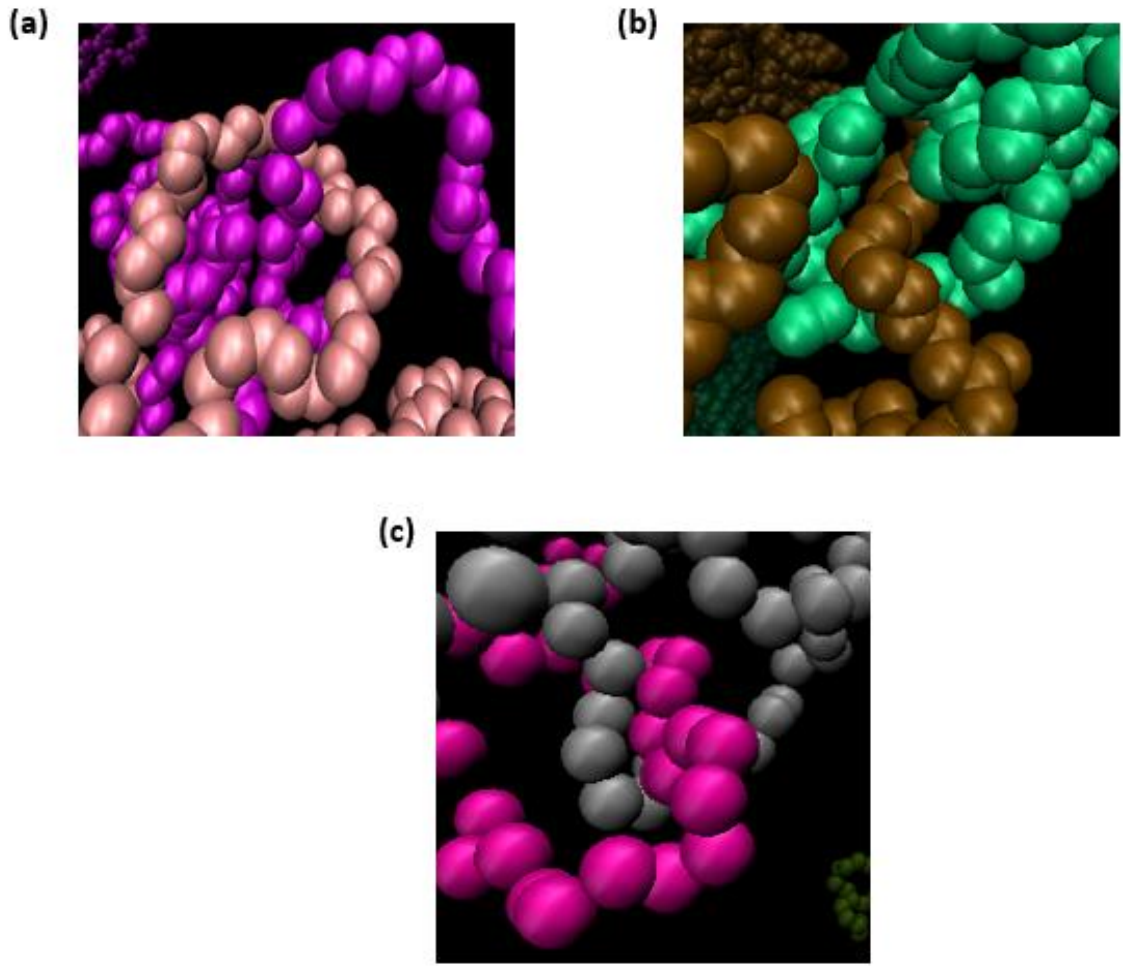

Figure 3. Entanglement points found after visually inspecting the system

Table 4: Conversion of physical quantities from reduced to SI units

| Quantity | Dimension | DPD Values | SI Units |
|---|---|---|---|
| Mass | M | 1 | $8.90 \times 10^{-26}$ Kg |
| Length | L | 1 | $7.36 \times 10^{-10}$ m |
| Energy | E | 1 | $4.142 \times 10^{-21}$ J. $K^{-1}$ |
| Timestep, dt | $L(M/E)^{1/2}$ | 0.01 | $3.41 \times 10^{-14}$ s |
| T | θ | 1 | 300 |

## 4. Results and Discussions

After the successful run of the DPD simulations, these simulations were analysed and various parameters like $n$, $N$, $v_e$ and $G_e$ were estimated and used for the modeling of the mechanical properties.

### 4.1 Estimation of $n$ and $N$

The parameters $n$ and $N$ are related to the Arruda Boyce model as mentioned earlier. For estimating $n$ and $N$, the crosslinked polymer chain segments were analysed from the post DPD simulations. The parameter $N$ can be thought as the number of polymer chain segments between two chemical cross-links per unit volume while the parameter $n$ can be thought of as the number of equivalent random links per polymer segment between two chemical crosslinks [20]. However, some of the cross-linking happens such that there are some bonds created between the same polymer chain, i.e. a closed loop formed due to such crosslinking. These same-chain closed loops do not take part in the load bearing structural network and hence they are removed from our counts [20]. The calculated value of $N$ based on the 200 chains system was estimated to be around $8.55 \times 10^{25}$ m$^{-3}$. Each simulation box contained 1600 crosslinking beads, but out of these, only 820 crosslinkers (around 51.25%) could make bonds with different adjacent polymer chains, and hence only the effective crosslinking bonds created by these beads were taken into our estimation.

For estimating the parameter $n$, we need to calculate the number of random links between two crosslinking points. These random links can be taken as Kuhn lengths measured along the backbone of the polymer chain segment [20,39]. In this study, we have tried to take a SBR chain with same composition as in Ref [8], but with half of the molecular weight of their SBR chain for ease of making the initial configurations in PACKMOL. In this system since half of the molecular weight of the original chain was considered, the polymer chain can be depicted by 288 Kuhn segments instead of 576 Kuhn segments in Ref [8]. In this study it was calculated that there were 8.19 crosslinking points per chain, hence the parameter $n$ could be estimated to be 31.36.

### 4.2 Estimation of parameters $v_e$ and $G_e$

The parameters $v_e$ and $G_e$ are taken from the ETM model, $v_e$ being the number of entanglements per unit volume and $G_e$ being the entanglement modulus. To estimate the parameter $v_e$, the Z1 code of Kroger et al was utilized [40]. The Z1 code analysis on our system gave the value of entanglement length $N_e$, equal to 17.057 using the M-Coil estimator [40]. From this value of $N_e$, we can estimate $Z$ (number of entanglements per unit chain) as 49.95 from the relation Z= number of beads per unit chain/$N_e$. Thus, the value of $v_e$ and $G_e$ for our system came out be $2.973 \times 10^{26}$/m$^3$ and 0.615 MPa, respectively.

## 4.3 Uniaxial Stress-Strain curve

Using the parameters estimated in the above sections and equation 5, a uniaxial stress-strain curve was plotted as shown in Figure 4. This estimated uniaxial curve was also compared with the experimental curve for the SBR rubber with same 1.75 phr of crosslinkers [27]. This experimental curve was also fitted against the CAM equation to give model-fit parameters $n$, $N$ and $G_e$. The fitting was done through the least squares minimisation techniques. The obtained model-fit parameters were $n = 26.99$, $N = 7.98 \times 10^{25}/m^3$ and $G_e = 0.412$ MPa which is very close to our estimated parameters through simulations ($n = 31.36$, $N = 8.55 \times 10^{25}/m^3$ and $G_e = 0.615$ MPa). A yellow colored stress-strain curve was drawn in Fig. 4 with the obtained model-fit parameters and equation 5. Since in this study the stress response due to cross-linking and entanglement is independent, another curve in blue is also shown for the relative effect of entanglement from the DPD based analysis on the total strength of the rubbery material in the figure. Thus, Fig.4 has four curves namely, the experimental curve, model-fit parameter obtained curve, multiscale model predicted curve and entanglement effect curves superimposed on each other.

It is evident from Fig. 4 that the multiscale model predicted curve matches very well with the experimental curve with $R^2$ value of 0.9836. This shows that our constitutive model combining the Arruda Boyce model and the Extended Tube model when coupled with multiscale DPD based molecular simulations can help in estimating the mechanical properties of the elastomers with great accuracy. However, the effect of entanglement is significant as the stress from the entanglement part was estimated to be around 50% of the total measured stress for a strain up to 150%. The relative effect of entanglement falls below 10% of the measured experimental stress only after undergoing a strain more than 550% for the studied rubber. This is in line with our previous paper [8] to model the hyperelastic behavior of this SBR material, where only crosslinking effect was considered and it was found that crosslinking alone underpredicted the stress for up to 150% strain. But by putting together both the crosslinking and entanglement effects in this work, the stress-strain curve from the DPD study closely matches the experimental data for the whole range of 600% stretch in Fig. 4. The differences of the predicted stress from DPD modeling and experimental values for stretches of 50%, 100%, 200%, 300%, 400%, 500% and 600% are only within 10%, 20%, 20%, 13%, 8%, 2% and 8%, respectively. This developed multiscale and multi-method framework can help in design and explore polymer matrix materials in an efficient manner. This type of discussed methodology will be helpful in reducing costly and time-consuming experiments on properties of polymeric materials.

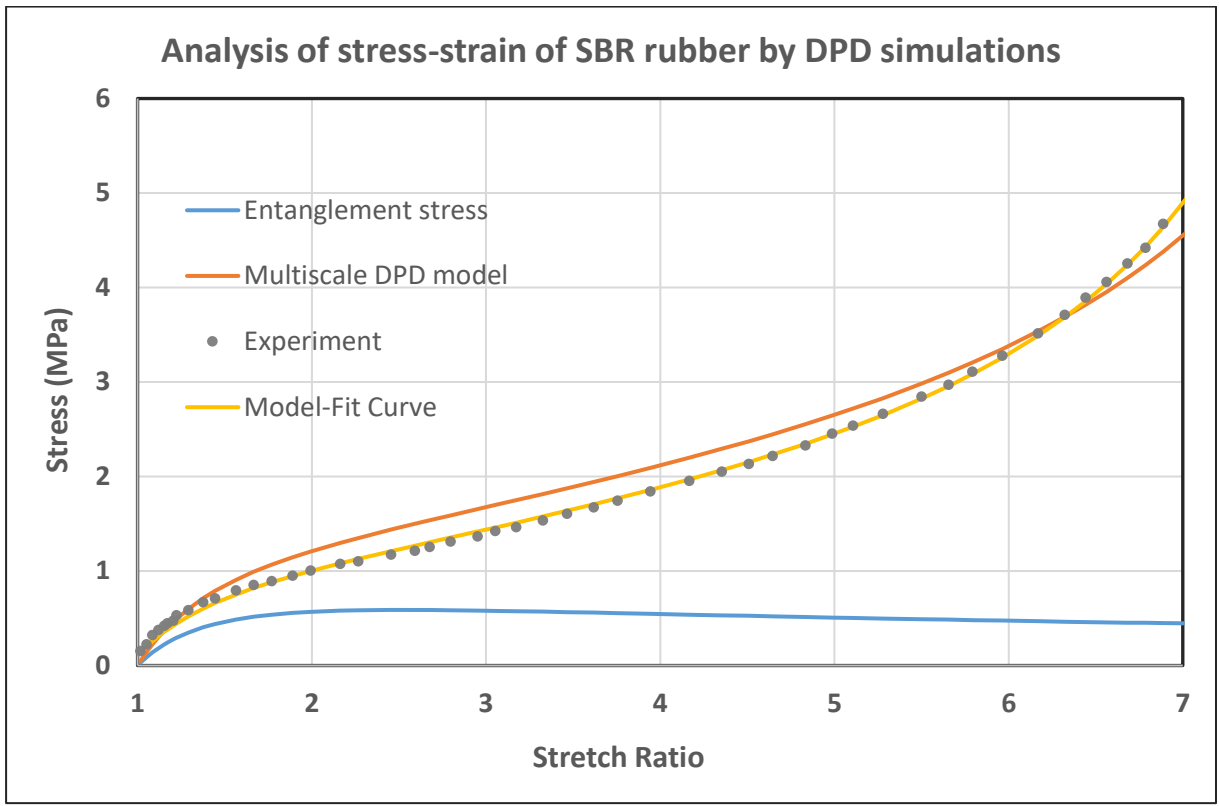

Figure 4. Different uniaxial stress-stretch curves from experiment, DPD multiscale simulation and CAM fitting. Effect of entanglement is shown along with the stress-strain in blue.

**Conclusion**

From the DPD based multiscale modelling study of the hyperelastic SBR rubber the following points can be concluded.

1. DPD based coarse grained simulations in combination with entanglement estimator and segmental repulsive parameter can give realistic estimate of entanglement density of long polymeric chain systems.
2. Cross-linking by DPD methods along with the calculated entanglement density can be utilized by constitutive analytical methods to predict accurate stress-strain relations of elastomers.
3. For lower levels of strain of 150% in rubbery elastomers, the effect of entanglement can be up to even 50% of the measured stress. However, for larger strains the relative effects of entanglement fall off.
4. Multiscale modelling based interdependent and connected all atomistic MD, DPD and CAM together has the capability to accurately predict the hyperelastic mechanical properties by modelling effective crosslinking and entanglement effects.